\documentclass[usenatbib,useAMS]{jaa}
\usepackage{graphicx}
\usepackage{natbib}
\usepackage{amssymb}
\usepackage{xcolor}
\usepackage{url}
\usepackage{float}

\newcommand\swift{{\it Swift}}

\newcommand\kms{\ifmmode {\rm~km\ s}$^{-1}$ \else ~km s$^{-1}$\fi}
\newcommand\Hunit{\ifmmode {\rm~km\ s}$^{-1}$\ {\rm Mpc}$^{-1}$
        \else ~km s$^{-1}$ Mpc$^{-1}$\fi}
\newcommand\ctssec{\ifmmode {\rm~count\ s}$^{-1}$ \else ~count s$^{-1}$\fi}
\newcommand\ergsec{\ifmmode {\rm~erg\ s}$^{-1}$ \else
        ~erg s$^{-1}$\fi}
\newcommand\funit{\ifmmode {\rm~erg\ s}$^{-1}$\ ; {\rm cm}$^{-2}$ \else
        ~ergs s$^{-1}$ cm$^{-2}$\fi}
\newcommand\phflux{\ifmmode {\rm~photon\ s}$^{-1}$\  ; {\rm cm}$^{-2}$
        \else   ~photon s$^{-1}$ cm$^{-2}$\fi}
\newcommand\efluxA{\ifmmode {\rm~erg\ s}$^{-1}$\ ; {\rm cm}$^{-2}$\ ; {\rm
        \AA}$^{-1}$ \else ~erg s$^{-1}$ cm$^{-2}$ \AA$^{-1}$\fi}
\newcommand\efluxHz{\ifmmode {\rm~erg\ s}$^{-1}$\ ; {\rm cm}$^{-2}$\ ; {\rm
        Hz}$^{-1}$ \else ~erg s$^{-1}$ cm$^{-2}$ Hz$^{-1}$\fi}
\newcommand\cc{\ifmmode {\rm~cm}$^{-3}$ \else cm$^{-3}$\fi}
\newcommand\FWHM{\ifmmode {\rm~FWHM} \else ${\rm~FWHM}$\fi}
\newcommand\Msun{\ifmmode M_{\odot} \else $M_{\odot}$\fi}
\newcommand\Lsun{\ifmmode L_{\odot} \else $L_{\odot}$\fi}

\newcommand\hbeta{\ifmmode {\rm H}\beta \else H$\beta$\fi}
\newcommand\Kalpha{\ifmmode {\rm K}\alpha \else K$\alpha$\fi}
\newcommand\nh{\ifmmode N_{\rm H} \else N$_{\rm H}$\fi}
\makeatletter

\newcommand{\Rmnum}[1]{\expandafter\@slowromancap\romannumeral #1@}
\makeatother
\usepackage{float}

\def\Msun{\mbox{M$_\odot$}}

\def\kms{{\rm\,km\,s^{-1}}}

\def\mathnew{\mathsurround=0pt}   
\def\simov#1#2{\lower .5pt\vbox{\baselineskip0pt  
\lineskip-.5pt\ialign{$\mathnew#1\hfil##\hfil$\crcr#2\crcr\sim\crcr}}}

\def\'#1{\ifx#1\i{\accent"13\i}\else{\accent"13#1}\fi}

\def\ie{i.e.}

\def\bf{\textbf{}}

\begin{document}\sloppy

%%paper title
%%For line breaks \\ can be used within title 
\title{Spectro-Timing Analysis of a highly variable narrow-line Seyfert 1 galaxy NGC~4748 with {\it AstroSat} and {\it XMM-Newton} }

\author{Main Pal\textsuperscript{1}, Neeraj Kumari\textsuperscript{2,3}, P. Kushwaha\textsuperscript{4}, K. P. Singh\textsuperscript{5}, Alok C. Gupta\textsuperscript{4}, Sachindra Naik\textsuperscript{2}, G. C. Dewangan\textsuperscript{6}, P. Tripathi\textsuperscript{6}, Rathin Adhikari\textsuperscript{1}, O. Adegoke\textsuperscript{7}, H. Nandan\textsuperscript{8}}
\affilOne{\textsuperscript{1}Centre for Theoretical Physics, Jamia Millia Islamia< New Delhi, India.\\}
\affilTwo{\textsuperscript{2}Astronomy and Astrophysics Division, Physical Research Laboratory, Ahmedabad, Gujarat, India.\\}
\affilThree{\textsuperscript{3}Indian Institute of Technology, Gandhinagar, Gujarat, India\\}
\affilFour{\textsuperscript{4}Aryabhatta Research Institute of Observational Sciences, Manora Peak, Nainital, India.\\}
\affilFive{\textsuperscript{5}Indian Institute of Science Education and Research, Mohali, India.\\}
\affilSix{\textsuperscript{6}Inter University Centre for Astronomy and Astrophysics, Pune, India.\\}
\affilSeven{\textsuperscript{7}Department of Astronomy, University of Geneva, Switzerland.\\}
\affilEight{\textsuperscript{8} Department of Physics, Gurukul Kangdi University, Haridwar, India.\\}

\twocolumn[{

\maketitle

\corres{rajanmainpal@gmail.com}

\msinfo{: 7 November 2020}{: 12 January, 2021.}

\begin{abstract}
We present a detailed timing and spectral study of an extremely variable narrow-line Seyfert~1 galaxy NGC~4748 using  observations in the year 2017 and 2014 performed with {\it AstroSat} and {\it XMM-Newton}, respectively. Both observations show extremely variable soft and hard X-ray emission that are correlated with each other. In the 2014 data set, the source retains its general behaviour of ``softer when brighter" while the 2017 observation exhibits a ``harder when brighter" nature. Such changing behaviour is rare in AGNs and is usually observed in the black hole binary systems. The ``harder when brighter" is confirmed with the anti-correlation between the photon index and the 0.3-10 keV power-law flux. This suggests a possible change in the accretion mode from standard to the advection-dominated flow. Additionally, both the observations show soft X-ray excess below 2 keV over the power-law continuum. This excess was fitted with a single or multiple blackbody component(s). The origin of soft excess during the 2017 observation is likely due to the cool Comptonization as the photon index changes with time. On the other hand, the broad iron line and delayed UV emission during the 2014 observation strongly suggest that X-ray illumination onto the accretion disk and reflection and reprocessing play a significant role in this AGN.

\end{abstract}


}]

\doinum{12.3456/s78910-011-012-3}
\artcitid{\#\#\#\#}
\volnum{000}
\year{0000}
\pgrange{1--}
\setcounter{page}{1}
\lp{1}

\section{Introduction}

Radio quiet Active Galactic Nuclei (AGNs) are a sub-class of AGNs that lack the radio jet emission. The AGNs are powered by accretion of material onto the central supermassive black hole (SMBH) through the accretion disk \citep{1963Natur.197..533H,1969Natur.223..690L}. These objects are found to be extremely variable over a wide range of time scales from minutes to years \citep{2003A&AT...22..661G, 2016ApJS..225...29B}. The variability properties such as amplitude and radiated power vary over different bands. The variability seen in the optical/Ultraviolet (UV) to X-ray bands has been studied extensively in several AGNs \citep{1990ApJ...359...86E, 2011fxts.confE..67R, 2007ApJ...668L.111G}. Highly variable X-ray emission is considered to be originated from the compact hot plasma while the UV/optical emission are considered to be originated from the accretion disk. Variations of various physical properties \ie~luminosity, accretion rate, and derived parameters such as the spectral indices obtained from the analysis of UV/optical and X-ray emission have been useful to understand these bright objects \citep{2019MNRAS.485.4790S,2019ApJS..241...33L,2006MNRAS.368..479G,2018MNRAS.473.3584P}. 

Sometimes, variations in the multi-waveband light curves, for example, in the X-ray to UV/optical bands, follow each other. The timing information of these changes suggest a common driving mechanism for their cause of origin. In several objects, the soft photons (\ie~,soft X-ray and UV/Optical photons) are observed before the hard X-ray photons. This is generally referred as the ``hard delay" or ``hard lag". The hard lag was observed first time in the binary black hole candidate GX~339-4 \citep{1991ApJ...383..784M, 1988Natur.336..450M}. After the discovery of hard lag in GX~339-4, similar type of delay was reported in a Seyfert~1 galaxy NGC~7469 by \citet{2001ApJ...554L.133P}. Since then, these lags have been detected in a number of AGNs \citep{2011MNRAS.416L..94E, 2004MNRAS.348..783M, 2019MNRAS.485.1454P}. This hard lag is generally interpreted in two ways: the light travel time resulting from single or multiple inverse Compton scattering in the hot plasma \citep{2020MNRAS.494.5056L, 2014MNRAS.439.1548A} and time taken by the density fluctuations of the accretion flow in the disk towards the centre \citep{2018MNRAS.476..225L}. In general, the time taken in the propagation of density fluctuations is much longer than the light travel time. Contrary, when the soft photons are detected after the hard X-ray photons, this delay is termed as the ``soft lag" or the``reverberation lag" \citep{2020MNRAS.492.1135V,2014MNRAS.439L..26K, 2011MNRAS.412...59Z, 2009Natur.459..540F}.

The correlated variations in the X-ray and the UV/Optical emission have been seen in a few AGNs. The changes in the UV/optical emission are sometimes observed before the variations seen X-ray emission. This type of process is expected when the UV/optical emission from the accretion disk are inverse-Comptonized in the plasma to give the power-law X-ray continuum. The time delay in X-ray emission is measured as the light crossing time scale or light travel time \citep{2008MNRAS.389.1479A,2019ApJ...870L..13A}. However, the fluctuations travel with the sound speed in the matter and thus hard lag has been seen over days to months and years time scales \citep{2008ApJ...677..880M, 2011A&A...534A..39M}. Sometimes, the variations in X-ray emission lead the changes in the UV/Optical emission. This is expected when X-ray emission from the hot plasma illuminates the accretion disk and get absorbed, it is re-radiated at longer wavelengths for example, in the UV/optical bands. This phenomenon is called X-ray reprocessing. This type of phenomenon has been observed in several different types of AGNs \citep{10.1093/mnras/stu1636, 2017MNRAS.466.1777P,2018MNRAS.474.5351P,2019BSRSL..88..143N, 2020MNRAS.498.5399H, 2020ApJ...890...47P}.   

Moderately correlated variations in X-ray and UV/Optical bands have also been reported and explored. There are several cases where the UV/optical emission and X-ray emission are found to be uncorrelated. These erratic flux variations in the UV/optical emission may be associated to the local anisotropic magnetic origin in the disk or the complex absorption/extinction present along the line of sight or the multiple mechanisms at the same time \citep{10.1093/mnras/stx2163, 1998ApJ...505..594N,2008RMxAC..32....1G, 2002AJ....124.1988M}. In such cases, the understanding of spectral energy distribution (SED) plays a crucial role. Though, the SED in X-ray to UV/optical band is complex due to the mixture of intriguing spectral components. The main components are (i) power-law continuum obtained from the inverse Comptonization of seed photons from the disk, (ii) the UV/optical continuum originated from the accretion disk, (iii) soft X-ray excess over the power-law continuum below $\sim$2 keV. The origin of the soft X-ray excess is not well understood to date. The possible explanation for this component is associated either with the cool Comptonization or the blurred reflection from partially ionized accretion disk \citep{2012MNRAS.420.1848D,2006MNRAS.365.1067C, 2016MNRAS.457..875P, Dewangan_2007}. 

NGC~4748 is a narrow-line Seyfert~1 galaxy \citep{2009ApJ...705..199B} located at a redshift of $z=0.014$. It harbours one of the lowest mass SMBH  ($M_{BH}=2.6_{-1}^{+1.6}\times10^{6}~M_{\odot}$) among AGNs  which makes it interesting and important in terms of its observed variability time scales. This nearby AGN has been found to be highly variable on short and long timescale as seen by {\it XMM-Newton} and {\it Swift} missions \citep{2018Ap&SS.363..228V, 2017BTSNU..56...18F}. \cite{2017BTSNU..56...18F} used {\it XMM-Newton} EPIC-pn data to investigate the short term variability and a $\sim10$ks periodicity was found in the 68ks long stare time. Using this periodicity, the upper limit on the mass of the blackhole was estimated to be $M_{BH}\sim 6\times10^{7}~M_{\odot}$. \citet{2018Ap&SS.363..228V} investigated the origin of the soft X-ray excess in the broad-band SED (0.5-500 keV range) using multiple mission archival data. Both the cool Comptonization and blurred reflection models were tested and found equally probable. In this work, we explore the variable nature of this AGN as well as the origin of the soft X-ray excess using {\it AstroSat} and {\it XMM-Newton} data. 

In the very next section, we describe the data used in the present work and the reduction procedures. In Section~3, we explore both long-term and short-term variability properties. Spectral analysis is presented in Section~4. Finally, we discuss our results in Section~5. In this paper, we used the cosmological parameters $H_{0} = 67.04~{\rm km~s^{-1}~Mpc^{-1}}$, $\Omega_m = 0.3183$ and $\Omega_{\Lambda} = 0.6817$ \footnote{http://www.kempner.net/cosmic.php} to calculate the distance.

\begin{figure}[!ht]
\begin{center}
\includegraphics[scale=0.55, angle=90]{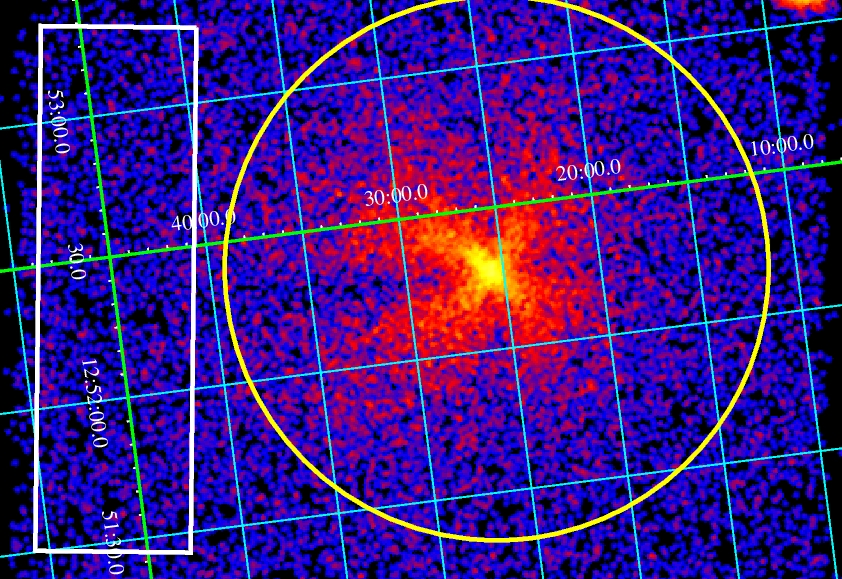}
\end{center}
\caption{Image of NGC~4748 from {\it AstroSat}/SXT data showing a circular region with radius of 14 arcmin centered at the source position and a rectangular box selected for source and background regions, respectively, for the extraction of light curves and spectra.}
\label{fig:fig-1}
\end{figure}

\begin{table*}
  \noindent \begin{centering}
    \caption{Details of the observations used in the present work. Start and end time are quoted in UTC.}
    \label{table:table-1}
    \begin{tabular}{llllc}

      \hline 
      Observatory & Obs.-ID      & Start time   & End time    & GTI (ks)\\
      \hline
      {\it AstroSat} &  9000001226  & 15 May 2017; 14:57:15 &17 May 2017; 14:27:36 & 56.8\\   
      {\it XMM-Newton$^{*}$} &0723100401    &14 Jan. 2014; 07:35:21 & 15 Jan. 2014; 01:45:03  &43.5\\
     \hline
     \end{tabular}
    \par\end{centering} {* : Optical Monitor (OM) observes filters UVW1, UVM2 and U partly overlapping to X-ray and X-rays were found to be flared affected.}
\end{table*}

\section{Observational data and its reduction}

We observed (PI: Main Pal) NGC~4748 with the Soft X-ray Telescope (SXT) onboard first Indian multi-wavelength space observatory {\it AstroSat} \citep{2014SPIE.9144E..1SS,AGRAWAL20062989}. The SXT consists of shells of conical mirrors that focus the X-ray photons in the 0.3--8 keV band onto a CCD detector. The field of view of the SXT is 40 arcmin. The effective area of the telescope is 90 cm$^2$  at 1.5 keV. The energy resolution of the detector is 90 eV at 1.5 keV and 136 eV at 5.9 keV. The detailed description of the SXT instrument is given in \citet{2016SPIE.9905E..1ES,2017JApA...38...29S}. Along with the SXT data, we also used publicly available archival data on NGC~4748 from the {\it XMM-Newton} mission. Details of the observations used are listed in Table~\ref{table:table-1}. Below, we describe the data reduction procedure followed in this work. 

\subsection{{\it AstroSat}/SXT}

In order to analyse the {\it AstroSat}/SXT data, we followed the standard analysis procedure as outlined in the {\it AstroSat} Science Support Cell webpage (ASSC\footnote{\url{http://astrosat-ssc.iucaa.in/}}). First of all, we obtained level~1 data from the {\it AstroSat} archive at the Indian Space Science Data Centre (ISSDC). We then run the {\tt sxtpipeline} tool to get clean event files calibrated with the latest calibration database provided by the instrument team. We merged all the event files obtained from each orbit of observation into a single event file. Using the {\tt xselect} task of {\tt FTOOLS}, we extracted scientific products from the merged event file. The source spectrum was extracted by selecting a circular region of 14 arc-minute radius centred at the source co-ordinates (see Figure~\ref{fig:fig-1}). The payload operation centre (POC) performed deeper observations to get good statistics for stable background spectrum of SXT instrument while background light curve extracted from our own observation shows some variations with lesser statistics. We therefore preferred background spectrum provided by the POC team over the background spectrum obtained from selected rectangular box near the source. The ARF was generated using latest merged event file and also corrected for the vignetting effect.

\begin{figure*}[!ht]
\begin{center}
\includegraphics[scale=1, angle=0]{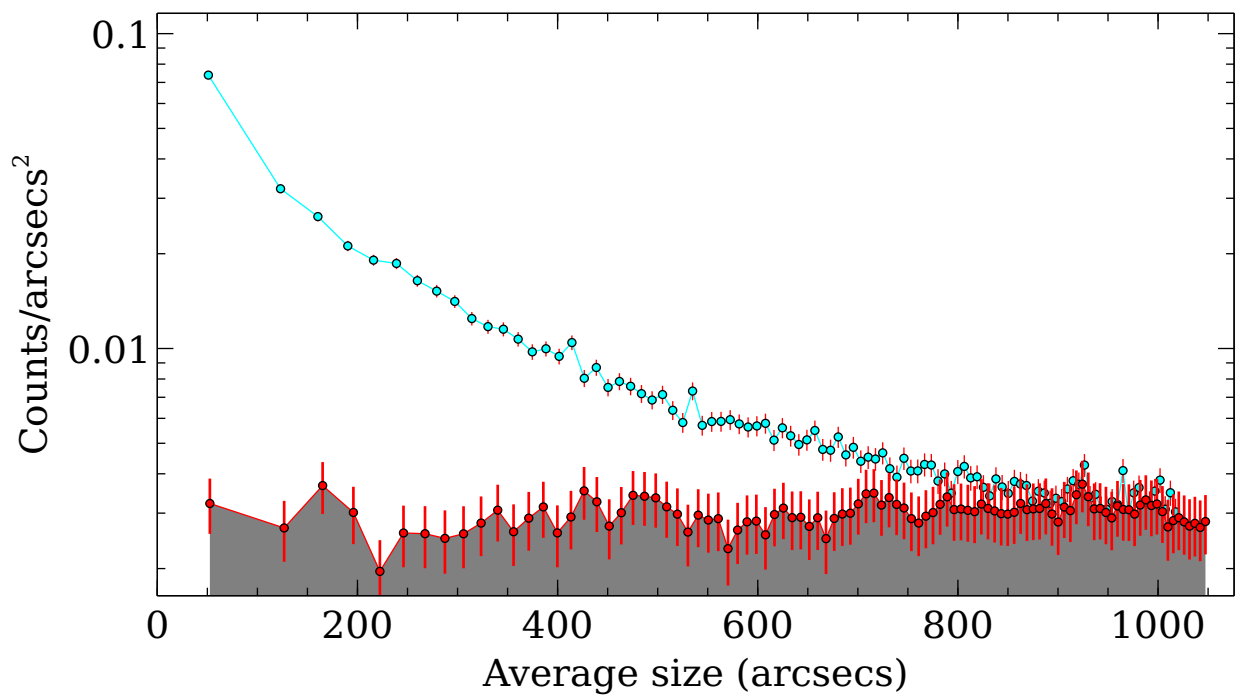}
\end{center}
\caption{{\it AstroSat}/SXT data: Radial profiles for source plus background (green) and background (red) are shown here for selected circular and rectangular regions in Figure~\ref{fig:fig-1}, respectively.}
\label{fig:fig-2}
\end{figure*}

To extract the source and background light curves, we checked the background light curve above 7 keV for any solar/proton flare and did not find any sharp change in the background light curve. We searched for a suitable region on the CCD of SXT to find the stable background. Fortunately, the observations of NGC~4748 were carried out in such a way that the image of the source is located at a marginally off-centered position on the CCD. We then estimated the radial profile of the source plus background down to 16.5 arcmin of a circular region. For background light curve, we used a rectangular region of length 27 arcmin and width 8 arcmin away from the source as shown in Figure~\ref{fig:fig-1}. The radial profiles of the selected source and background regions are shown in Figure~\ref{fig:fig-2}. Clearly, we found a stable background with marginal variations away from the source during the {\it AstroSat} observation of NGC~4748. This selected rectangular region was used for the extraction of only background light curves in different energy bands. We extracted light curves of background and source plus background in three energy ranges  namely 0.6-2 keV, 2-7 keV and 0.6-7 keV bands using {\tt xselect} task. We then obtained background subtracted light curves and then computed the hardness ratio for the source.

  \begin{table*}
 
    \caption{Best-fit parameters of NGC~4748 obtained from the spectral analysis of {\it AstroSat}/SXT and {\it XMM-Newton} EPIC-pn data.  Results from the time resolved spectroscopy analysis of {\it AstroSat}/SXT are also listed. Flux is measured in $10^{-12}\rm~ergs~s^{-1}~cm^{-2}$. ``T" represents the time segment of long SXT light curve , which is used to extract the spectral parameters.}
    \label{table:table-2}
    \begin{tabular}{lllcccccc cccc}
      \hline 
      Model & Model  comp.     & SXT    & EPIC-pn & T1& T2&T3         \tabularnewline
    \hline              
        \\
  \underline{Tbabs}&     N$_{H}$~ ($10^{20}\rm~cm^{-2}$)       & 3.6 (fixed)  & 3.6 (fixed) & 3.6 (fixed) & 3.6 (fixed)    & 3.6 (fixed)       \tabularnewline   
  \underline{Power-law}&      $\Gamma$  &$2.07_{-0.07}^{+0.06}$  &$2.15_{-0.03}^{+0.02}$ &$1.96_{-0.10}^{+0.10}$  &$1.91_{-0.09}^{+0.08}$ & $2.67_{-0.10}^{+0.10}$    \tabularnewline
      &  Norm. $(10^{-3})$ & $2.81\pm0.13$ &$4.04_{-0.13}^{+0.08}$ & $3.5\pm0.3$ &$3.3_{-0.2}^{+0.2}$ & $1.55_{-0.06}^{+0.06}$   \tabularnewline
      &  F$_{0.3-2 keV}$ &$8.7\pm0.6$ &$12.8_{-0.4}^{+0.6}$ &$10.5\pm1.2$ & $9.9\pm0.8$  & 6.0$\pm0.4$  \tabularnewline
         %&F(2-10 keV) &6.56 &    \tabularnewline
      &  F$_{0.3-10 keV}$ &$15.24\pm0.36$ & $21.1^{+0.45}_{-0.30}$ &$20.1\pm0.8$ & $19.8\pm0.6$ &$7.5\pm0.3$    \tabularnewline
   % \\
      %  \underline{Blackbody-1}\\ 
    %\\
  \underline{Blackbody-1}&      $kT_{bb}$~\rm~(in eV)   &$120.6_{-10.3}^{+10.1}$ & $83.5\pm2.5$ &$133.9_{-14.2}^{+13.7}$ & $113.4\pm13.0$ & --  \tabularnewline
    &    Norm. $(10^{-5})$   &$6.0_{-1.0}^{+1.2}$ &$10.3_{-0.5}^{+0.8}$  &$8.2_{-1.6}^{+1.8}$ &$10.4_{-2.4}^{+3.7}$ &--   \tabularnewline
     &   F$_{0.3-2 keV}$ &$3.5\pm0.5$ & $3.95_{-0.30}^{+0.37}$ &$5.1\pm0.9$ &$5.7\pm1.1$& --  \tabularnewline
   % \\
       % \underline{Blackbody-2}\\
%    \\
    \underline{Blackbody-2}&    $kT_{bb}$~(in eV)  &--   &$178.8_{-9.2}^{+11.2}$&-&-&-  \tabularnewline
        &Norm. $(10^{-5})$&-- & $3.0_{-0.3}^{+0.4}$ &-&-&-  \tabularnewline
        &F$_{0.3-2 keV}$  &-- &$2.15\pm0.20$  &-&-&-   \tabularnewline
        % &F$_{net}$ (0.3-2 keV) &-- &6.1    \tabularnewline
 %   \\
       % \underline{Gaussian Line}\\
  %  \\
    \underline{Gaussian Line}&    E(keV) &  -- & $6.88_{-0.15}^{+0.13}$ &-&-&- \tabularnewline
    &    Width (keV) &-- & $\ge 0.77$  &-&-&-  \tabularnewline
     &   Norm. $(10^{-5})$ &-- & $5.2_{-1.3}^{+0.6}$ &-&-&-  \tabularnewline
      \hline
      Stats& $C/dof$   & 596.4/526       & 190.5/171 &-&1282.7/1295&- \tabularnewline

      \hline
    \end{tabular}
    %\par\end{centering} {Note--UVOT observes all optical/UV filters except IVth time series where only UVW2 filter was used.}
\end{table*}

\subsection{XMM-Newton}
NGC~4748 observation was performed with the European Photon Imaging Camera (EPIC) \citet{struder2001} and Optical Monitor (OM) \citep{2001A&A...365L..36M} onboard the {\it XMM-Newton} observatory in January 2014. The Science Analysis System (SAS 15.0 ; \citet{2004ASPC..314..759G}) software package was used for the data reduction. We extracted data from  the EPIC-pn and OM using the standard procedure described in the ``ABC Guide" of {\it XMM-Newton}. We processed the raw data with the latest calibration using the {\tt epproc} task and obtained the event files. The background light curve above 10 keV showed several solar/proton flares during the observation. We generated the good time intervals after removing the count rates above 1.1 count s$^{-1}$. After removing the flaring duration from the data, we got $\sim43$ ks of useful data. We used the single and double events (pattern $\le 4$) for EPIC-pn and ignored the events present at the edges and on bad pixels (flag=$0$). We selected a circular region of 50 arcsec centred at the source co-ordinates for the source spectrum and light curves. We extracted the background spectrum and light curves from a circular region of the same size located away from the source and free from any other source. The observation was not affected by pile-up effect significantly. We generated the response files such as redistribution matrices and auxiliary files using {\tt rmfgen} and {\tt arfgen} tasks, respectively. We grouped the spectra with a minimum of one count per bin with an oversampling factor of three using {\tt specgroup} {\tt SAS} tool. For the OM data, we extracted only UVW1 and UVM2 bands in fast mode using {\tt omfchain} tool.

\section{Temporal analysis}

\begin{figure*}[!ht]
\begin{center}
\includegraphics[scale=0.55, angle=0]{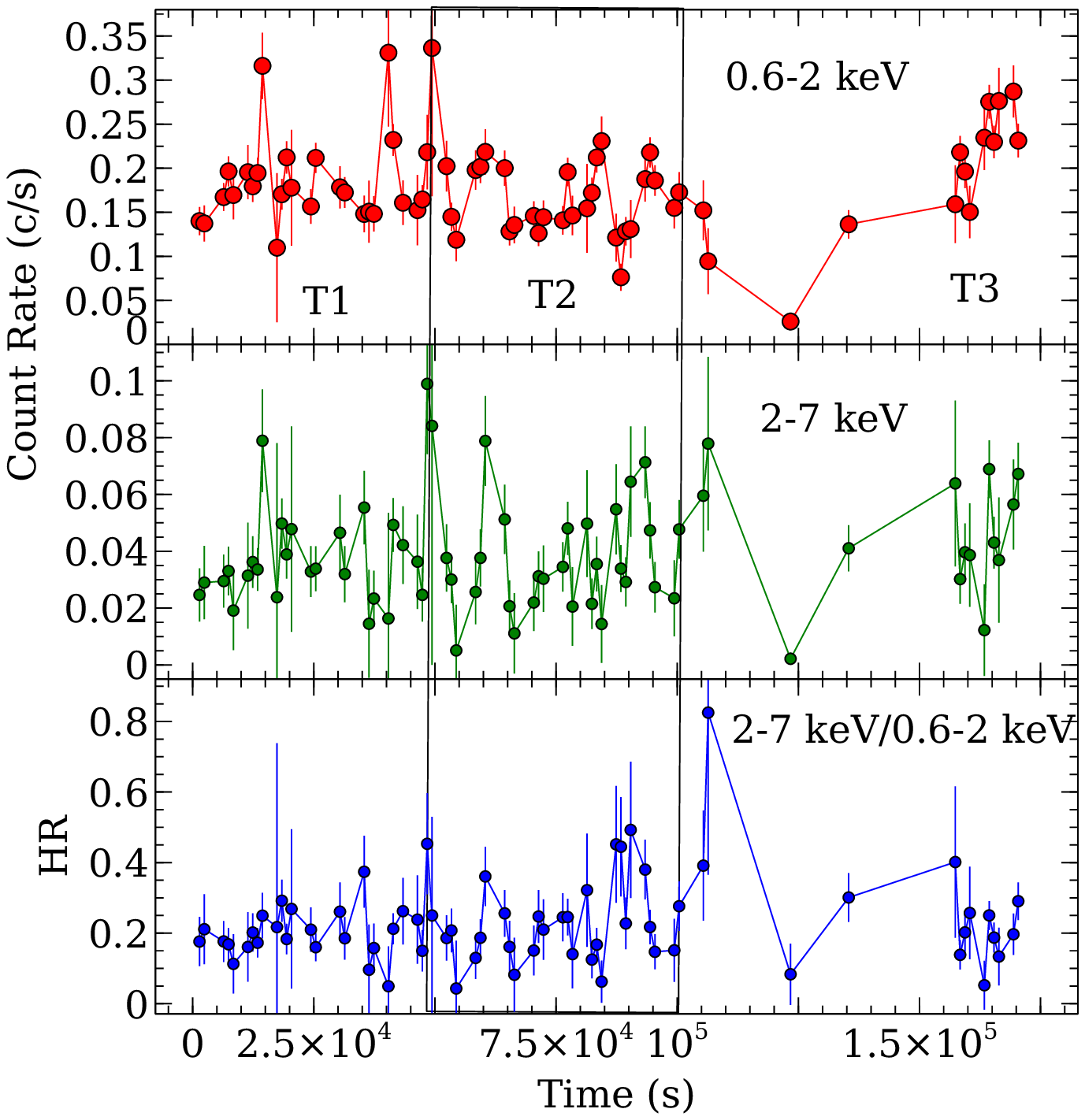}
\includegraphics[scale=0.55, angle=0]{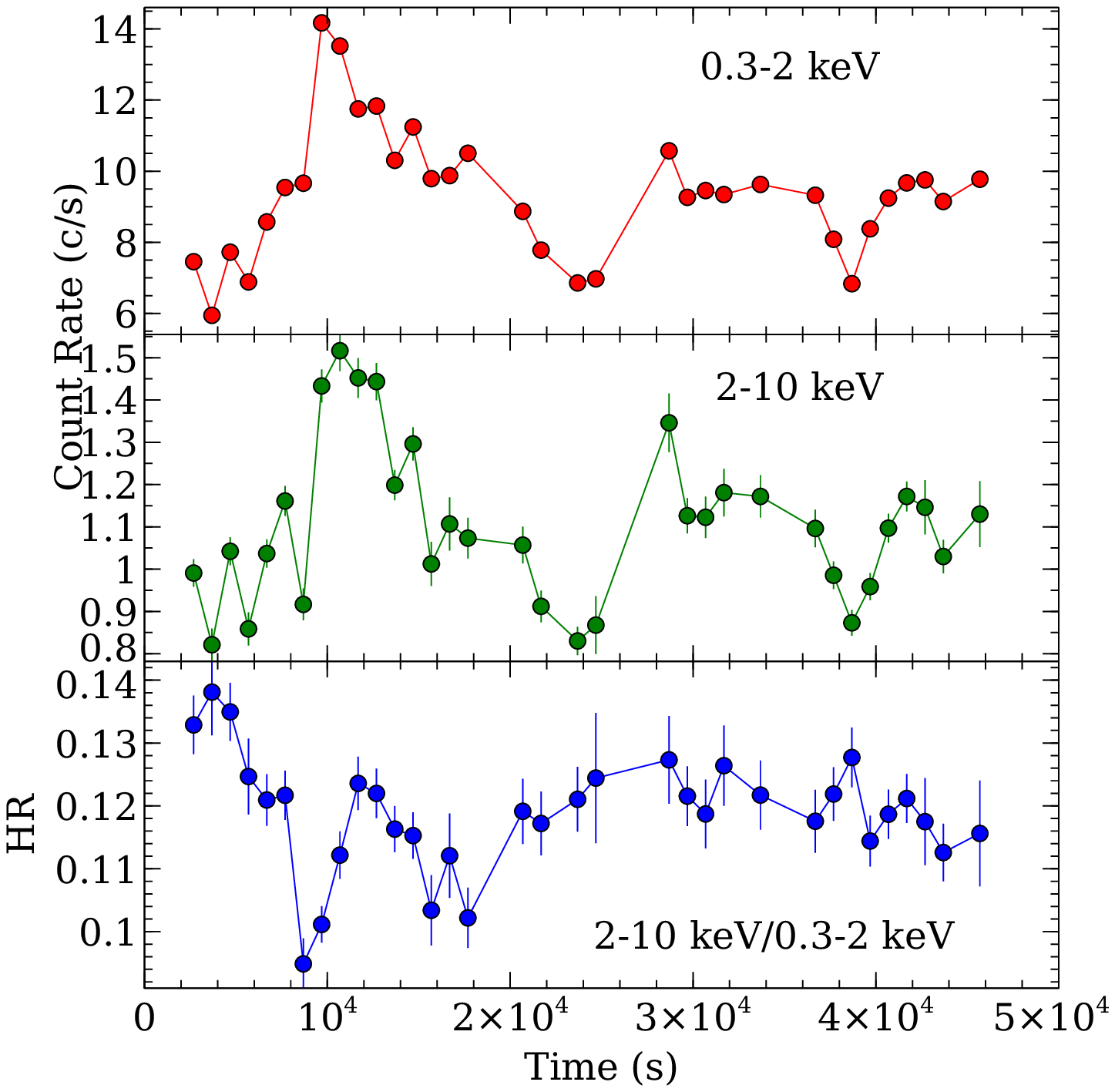}
\end{center}
\caption{ 1000s binned light curves in the soft and hard X-ray bands (top and middle panels) and hardness ratio (bottom panels) are shown here for \textbf{Left}: {\it AstroSat}/SXT data (soft X-ray band: 0.6-2 keV, hard X-ray band: 2-7 keV) and \textbf{Right}: {\it XMM-Newton} Epic-pn data (soft X-ray band: 0.3-2 keV, hard X-ray band: 2-10 keV). The time-binsize for each light curve is 1000 seconds. The vertical lines divide entire SXT light curve into three segments (T1, T2 and T3) which are used for time-resolved spectroscopy analysis.}
\label{fig:fig-3}
\end{figure*}

Soft (\ie~below 2 keV) and hard (\ie~above 2 keV) X-ray light curves obtained from the {\it AstroSat}/SXT and {\it XMM-Newton}/EPIC-pn data, binned at 1000~s, are shown in Figure~\ref{fig:fig-3}. The top and middle panels in left side show the background subtracted light curves obtained from the SXT data in the 0.6--2 keV and 2--7 keV bands, respectively. The hardness ratio i.e. the ratio between the 2--7 keV and 0.6--2 keV light curves is also shown in the left bottom panel of the figure. Variations based on binning time-interval and during the net exposure can be considered using ratio of the maximum to the minimum count rates/flux (${F_{max}/F_{min}}$). On visual inspection, we can see that the short term variability on 1000~s time scale is by about a factor of three, whereas the long term variability during the two days observation is found to be striking (by a factor of about thirteen) in the 0.6-2 keV band. In the 2-7 keV band, the changes on 1000~s time scale is found to be about a factor of four while the source is extremely variable (by a factor of forty five)  during the observation of about two days long used in this study. Different variability pattern suggests the presence of different spectral components in both the bands. Similarly, the hardness ratio shows large variation within the {\it AstroSat} observation and these changes seem interesting. 

The 2014 observation of NGC~4748 performed with EPIC-pn also showed variation in the soft X-ray (0.3-2 keV) and hard X-ray (2-10 keV) light curves (right panels of Figure~\ref{fig:fig-3}). The observed changes on binning time scale (1000~s) are about a factor of two and the variations within the net exposure are found to be more than a factor of two in the 0.3-2 keV band. In the hard band, we found a similar level of variability. Both the light curves seem correlated to each other. However, the variations in the hardness ratio show a nice pattern but opposite to that observed in the soft and hard X-ray bands. Therefore, the hardness ratio and its relation to the hard X-ray band would be crucial to investigate the cause of their variation. 

\begin{figure*}[!ht]
\begin{center}
\includegraphics[scale=0.6, angle=0]{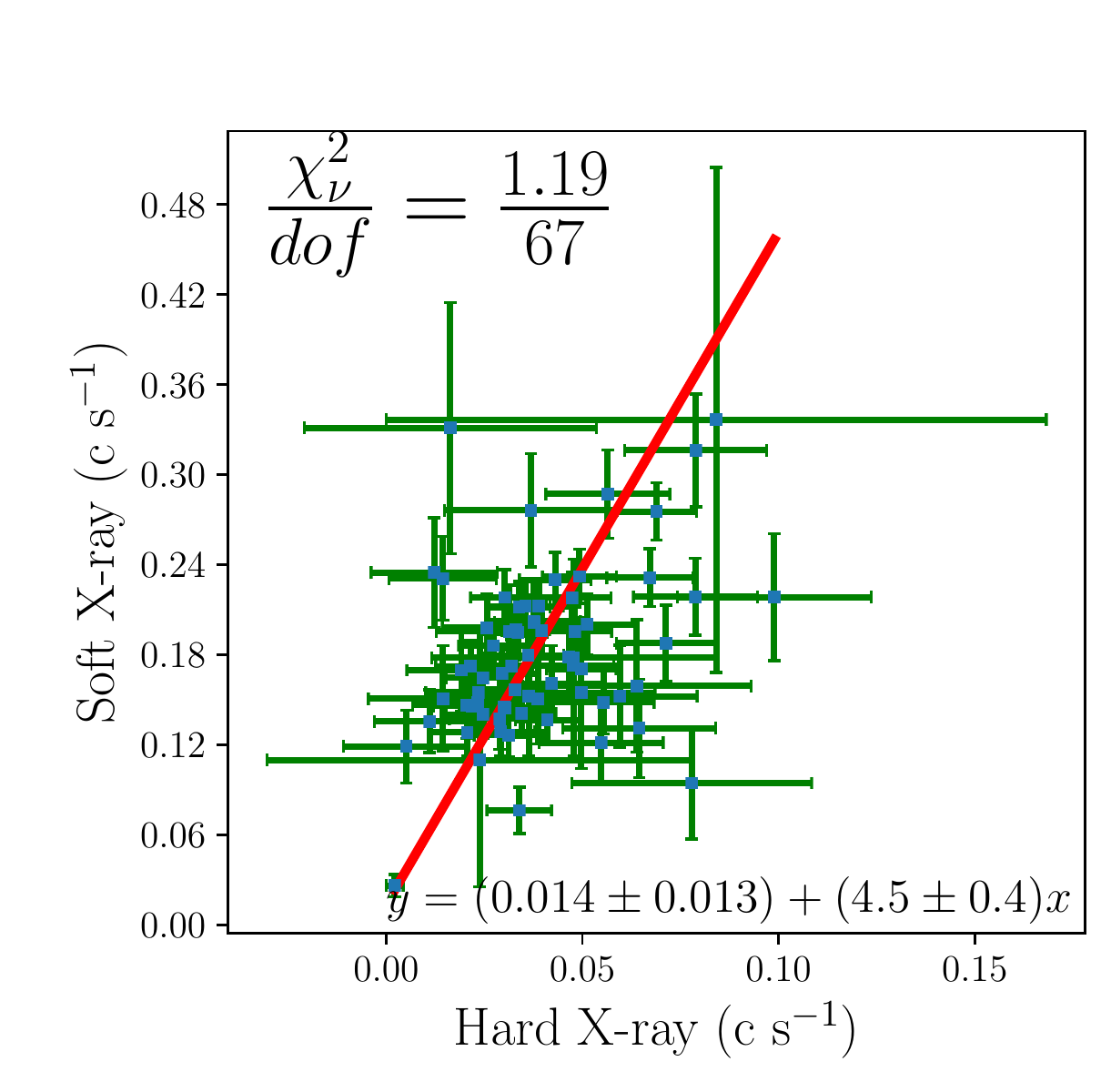}
\includegraphics[scale=0.5, angle=0]{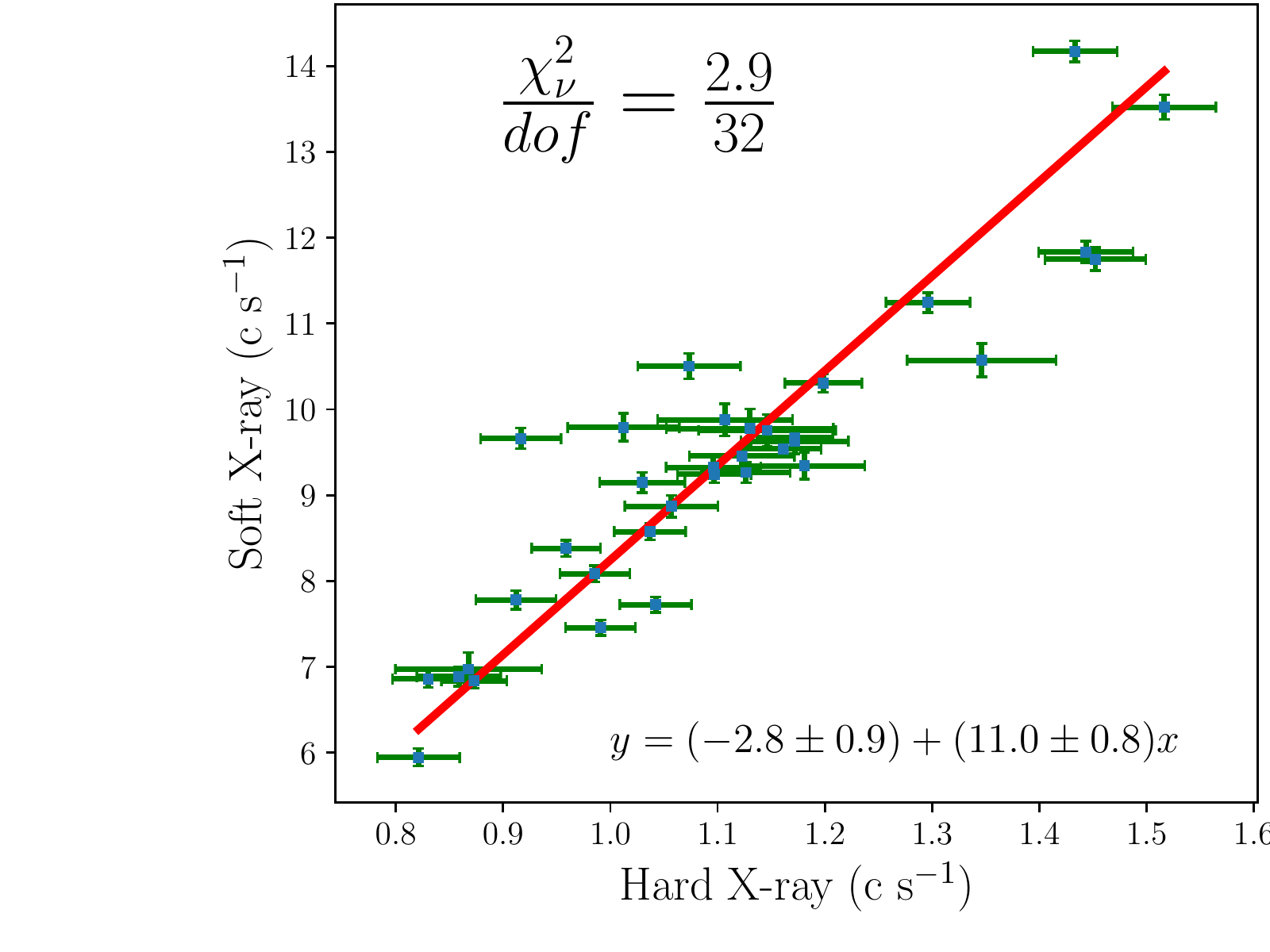}
\end{center}
\caption{The correlation between the soft and hard X-ray light curves are presented (\textbf{left panel}: SXT data and \textbf{right panel}: EPIC-pn data). The red line in each panel shows linear-fit function.}
\label{fig:fig-4}
\end{figure*}

The rapid variations present in each light curve of different energy bands appear to show a similar trend during both the observations. In order to understand the correlation between the soft and hard X-ray bands, we used the Pearson's correlation coefficient~`$\rho_{xy}$'. However, measuring~`$\rho_{xy}$' is not enough to relate the variables as the coefficient~`$\rho_{xy}$' does not take into account the two-dimensional normal distribution of both the variables. The probability $p$($\rho_{xy}$,n) is a useful quantity to determine the significance of correlation. This quantity determines how unlikely a given correlation coefficient $\rho_{xy}$ would be to yield no association between the variables of a sample. Small values (\ie $\le 0.05$) states that the observed variables are likely correlated. We determined the inter-band correlation coefficient and the probability (p-value) using the Python routine of the above technique. We found the correlation coefficient $\rho_{xy}=0.36$ and probability $p=0.002$ for soft and hard X-ray bands for the SXT data (left panel of Figure~\ref{fig:fig-4}). For EPIC-pn data, the corresponding values are $\rho_{xy}=0.92$ and $p=3.5\times10^{-14}$ for soft and hard X-ray bands and it is shown in the right panel of Figure~\ref{fig:fig-4}. From this analysis, it is clear that there exists some relationship between the soft and hard X-ray bands.

\begin{figure*}[!ht]
\begin{center}

\includegraphics[scale=0.7, angle=0]{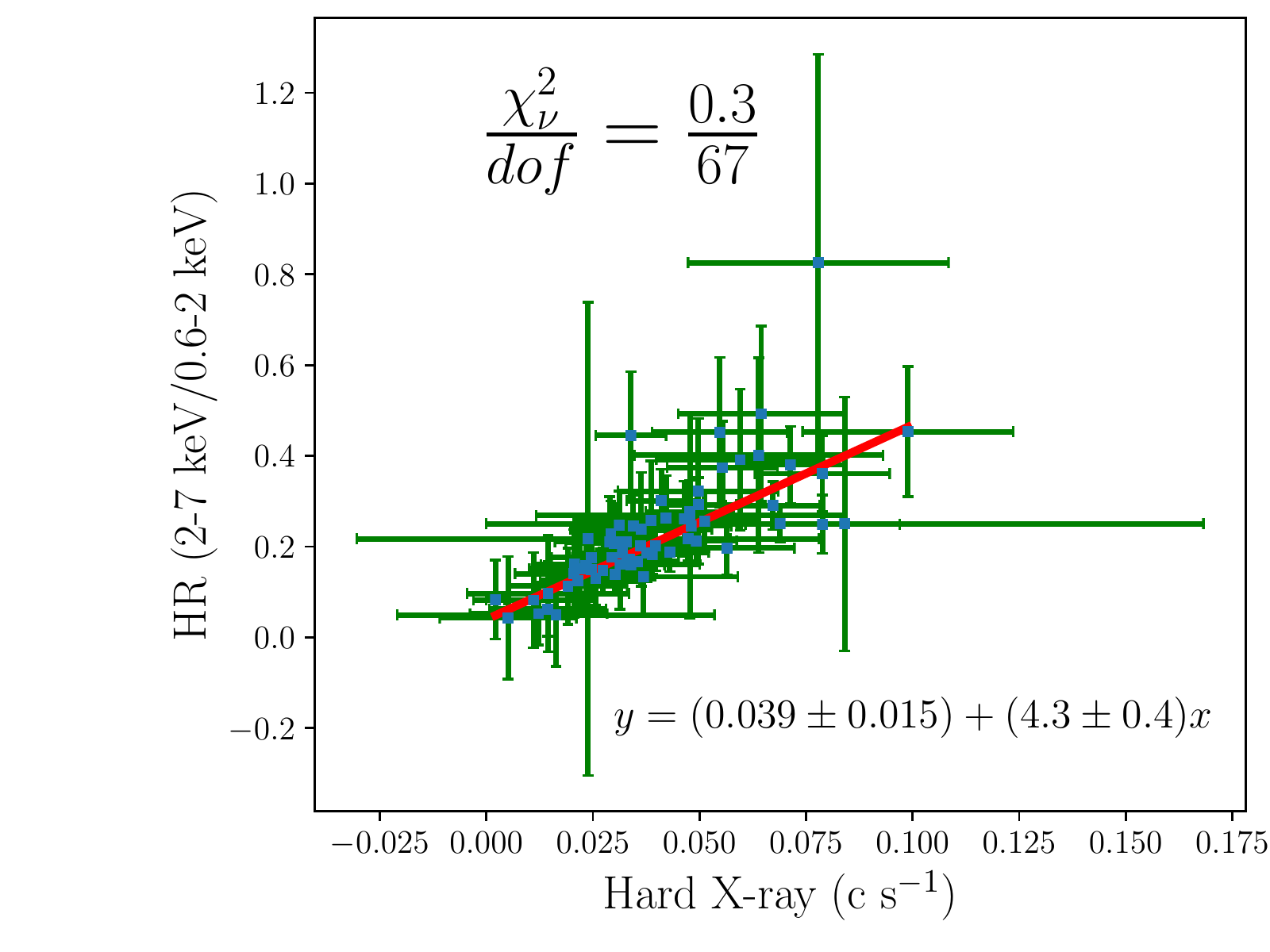}
\includegraphics[scale=0.56, angle=0]{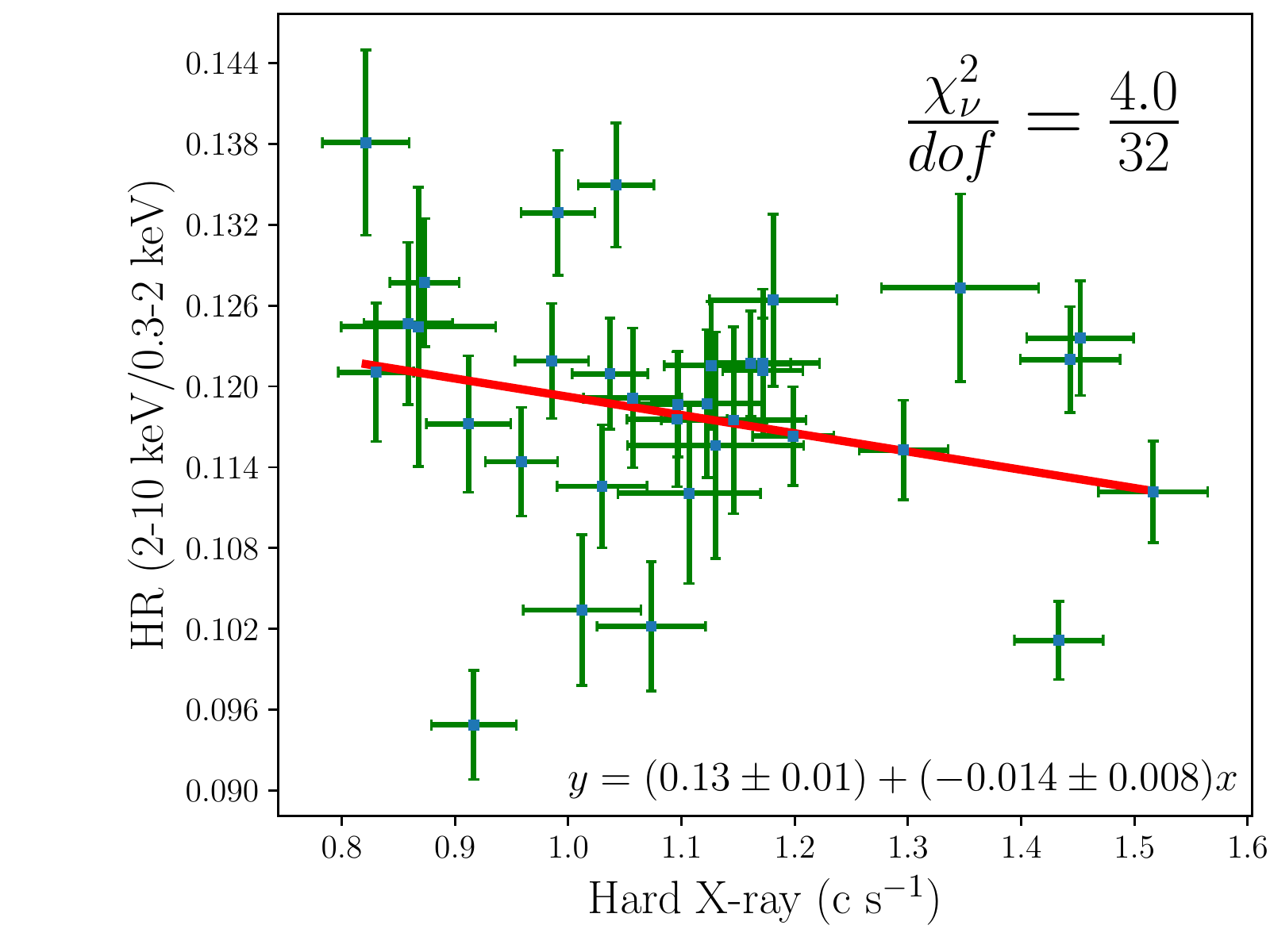}
\includegraphics[scale=0.56, angle=0]{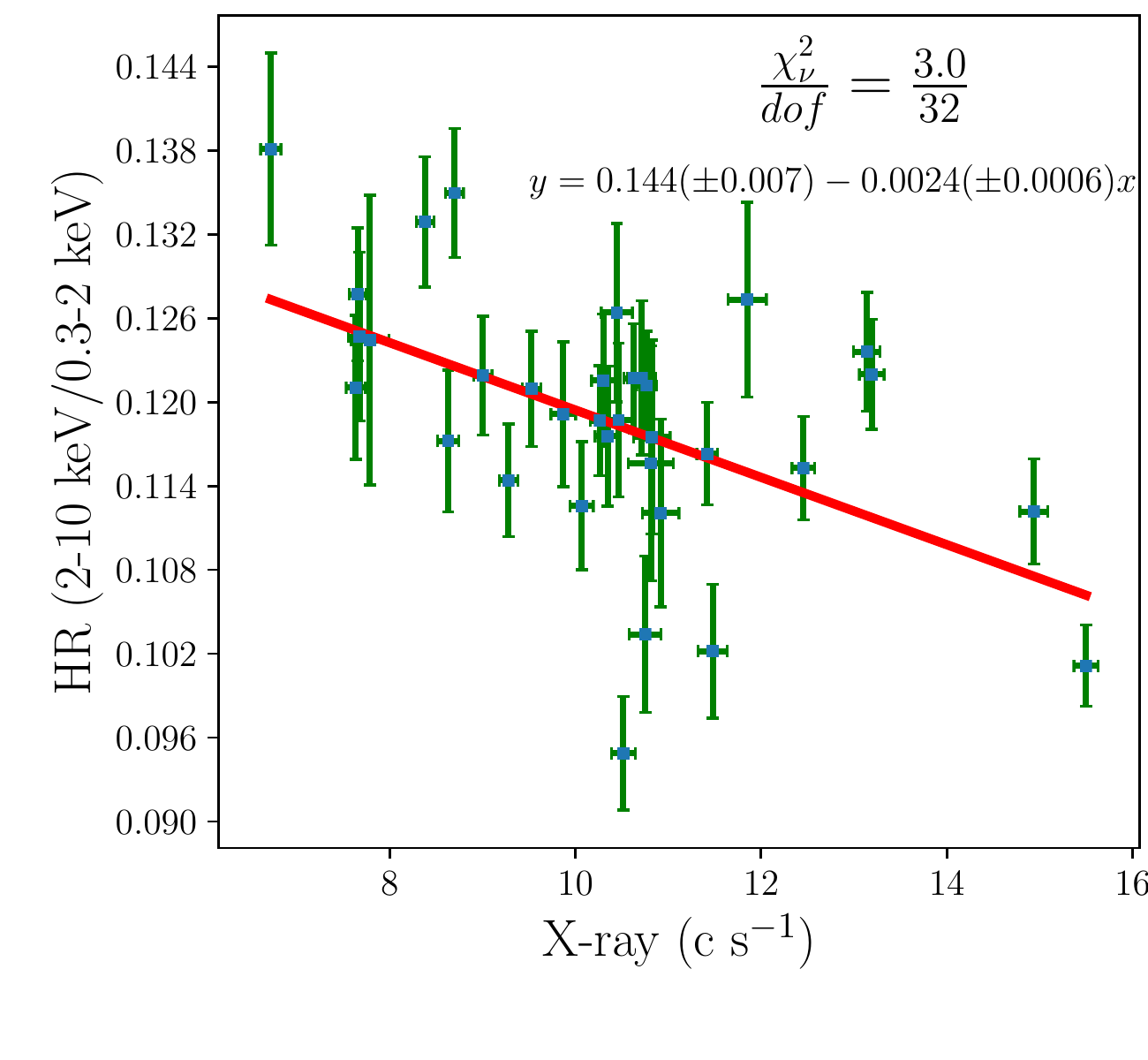}

\end{center}
\caption{Correlation between hard X-ray band and hardness ratio for SXT data (\textbf{top panel}) and EPIC-pn data (\textbf{bottom-left panel}). \textbf{Bottom-right}: correlation between hardness ratio and X-ray band (0.3-10 keV) of EPIC-pn data. The linear best-fit equation is shown with red colour in each panels of the figure. }
\label{fig:fig-5}
\end{figure*}

The observed hardness ratios are extremely variable during both 2014 and 2017 observations. The variations seen in the hardness ratio, sometimes, follow the hard X-ray band. We estimated the Pearson correlation coefficient $\rho_{xy}$ between the hardness ratio and hard X-ray band. We found $\rho_{xy}=0.75$ and $p=1.4\times10^{-13}$ for the hardness ratio and hard X-ray band for the SXT data, while for the EPIC-pn data, corresponding values are $\rho_{xy}=-0.55$ and $p=7.9\times10^{-4}$, respectively. The correlations are shown in the left and right panels of Figure~\ref{fig:fig-5} for SXT and EPIC-pn, respectively. Thus, highly variable soft and hard X-ray bands appear to be strongly correlated, however the hardness ratio with hard X-ray band is negatively correlated for EPIC-pn data (right panel of Figure~\ref{fig:fig-5}). The linear fit to the observed correlation also supports that the 2017 observation is positively correlated while the 2014 data exhibits negative correlation. To understand this contrary behaviour, we performed a systematic spectral analysis.

\section{Spectral analysis}

\subsection{Long term variability: average spectrum of 2014 and 2017 observations}

We employed C-statistic for finding the best fit model and to estimate 1$-\sigma$ range for each parameter.

\begin{figure*}[!ht]
\begin{center}
\includegraphics[scale=0.33, angle=-90]{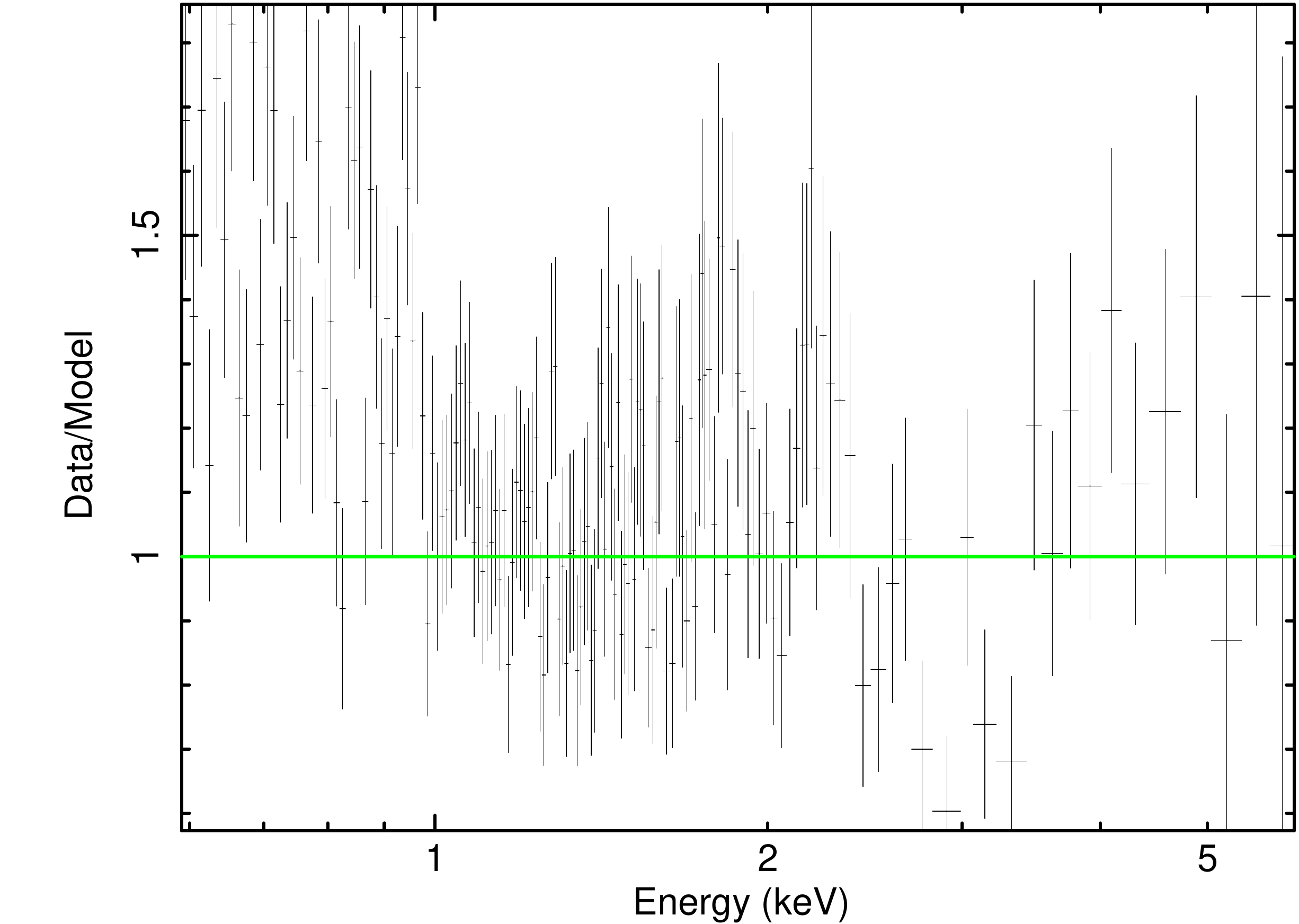}
\includegraphics[scale=0.33, angle=-90]{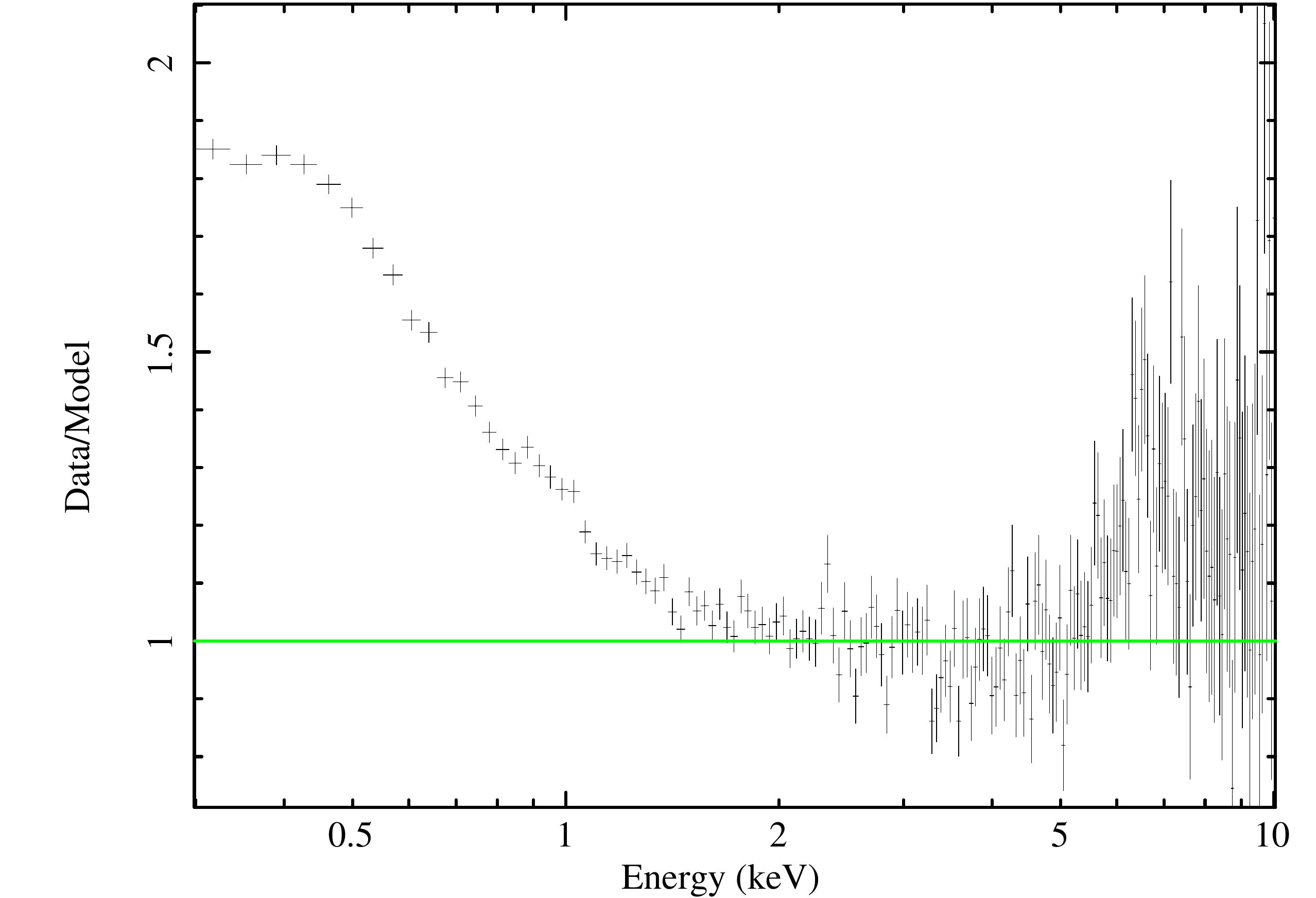}
\end{center}
\caption{\textbf{Left}: Absorbed {\tt powerlaw} model fitted in the 2-6 keV range and then extrapolated down to 0.6 keV for SXT data. A marginal excess is seen below 1 keV. \textbf{Right}: Absorbed {\tt powerlaw} model in the 2-5 keV band and then extrapolated down to 0.3 keV and up to 10 keV bands for EPIC-pn data. A strong soft X-ray excess below 2 keV and a broad emission line near 6 keV are found.}
\label{fig:fig-6}
\end{figure*}

\begin{figure*}[!ht]
\begin{center}
\includegraphics[scale=0.33, angle=-90]{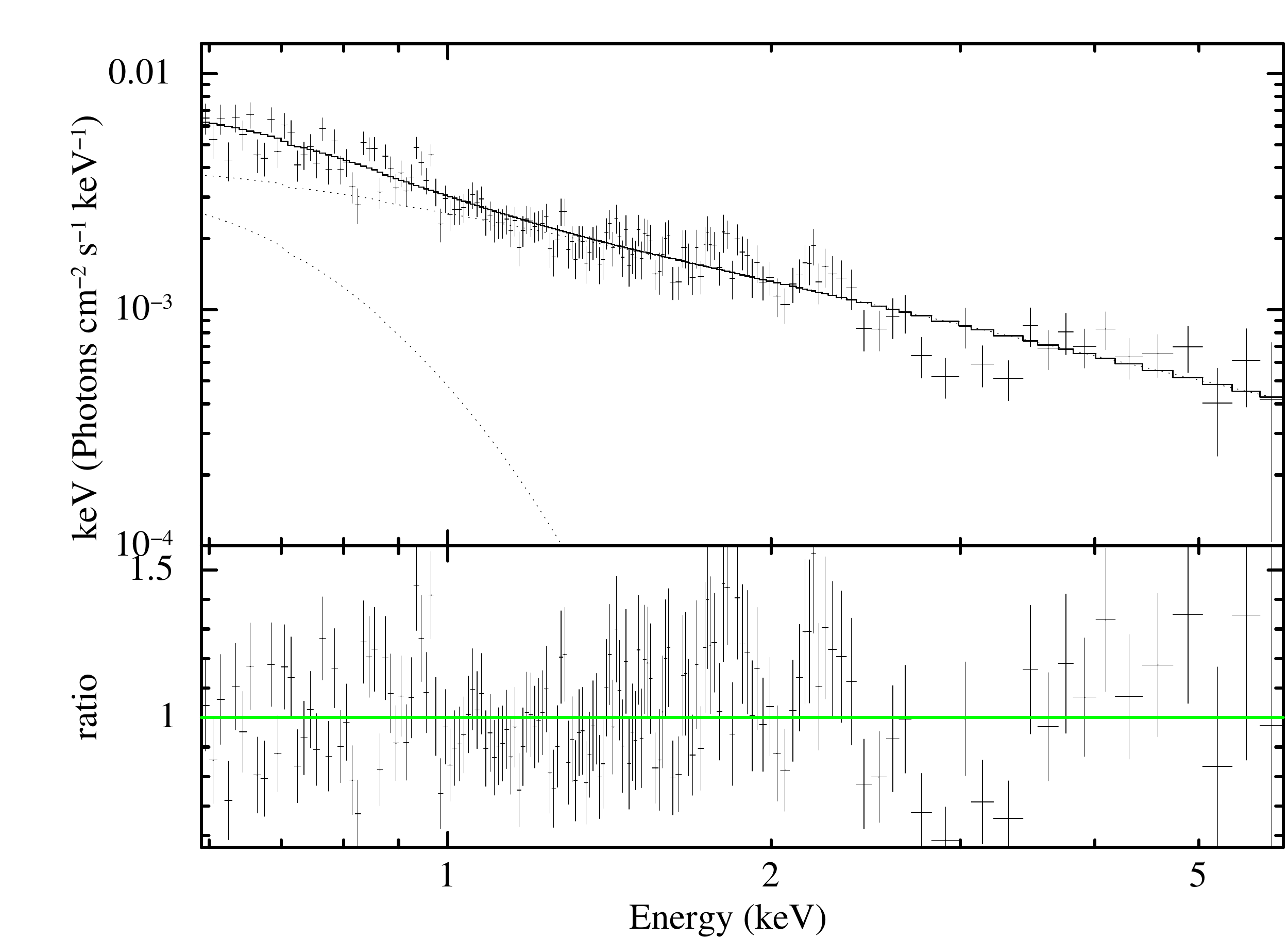}
\includegraphics[scale=0.33, angle=-90]{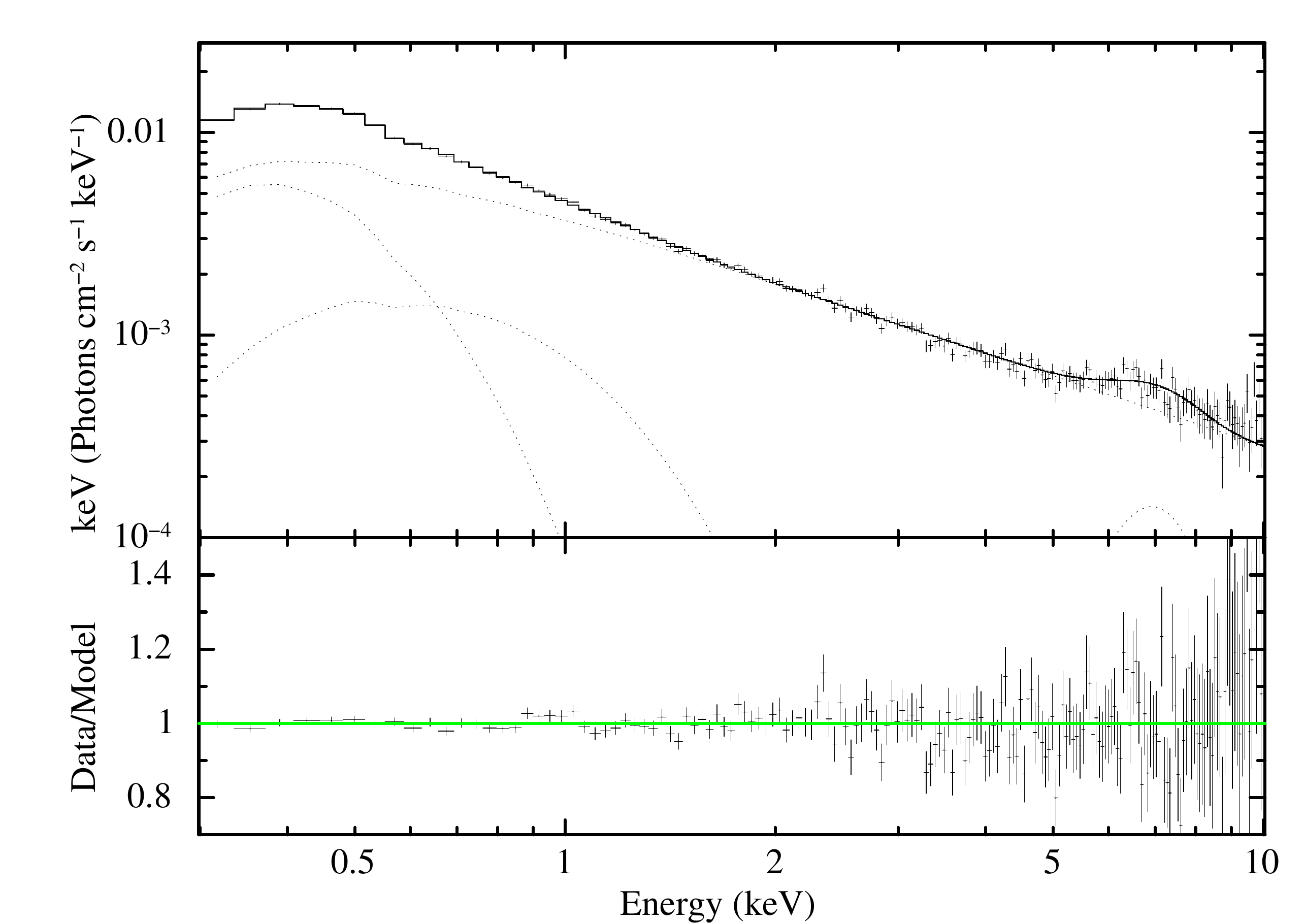}
\end{center}
\caption{ Best-fit model (solid line) consisting of an absorbed power-law (dotted line) and blackbody(s) (dotted curve) components, data and residuals (data/model) for {\it AstroSat}/SXT data in the 0.6-6 keV (\textbf{left panel}) and {\it XMM-Newton}/EPIC-pn data in the 0.3-10 keV band (\textbf{right panel}).}
\label{fig:fig-7}
\end{figure*}

We used the SXT data in the 0.6-6 keV range for spectral analysis. Below 0.6 keV, the response of the detector is uncertain, whereas above 6 keV, the data is dominated by background. In the beginning, the power-law continuum model {\tt powerlaw} was used to fit the data in 2-6 keV band. We then modified the model for the Galactic absorption by using {\tt tbabs} model and fixed it to the Galactic column density $N_{H}=3.6\times10^{20}~cm^{-2}$ \citep{2016A&A...594A.116H}. The best-fit power-law photon index in the 2-6 keV band was found to be $\Gamma=2.08\pm0.15$ with the fit statistics $C/dof=412.8/387$, where dof is the degree of freedom. Then the fitted band was extrapolated down to 0.6 keV and a marginal soft excess was found over the power-law continuum (see left panel of Figure~\ref{fig:fig-6}). To model this soft excess, a redshifted blackbody model {\tt zbbody} was used. The best-fit blackbody temperature was found to be $kT_{bb}\sim120$~eV. After freeing the index parameter for the best-fit model {\tt tbabs$\times$(powerlaw+zbbody)}, the fit statistic resulted in  $C/dof=596.4/526$. The best-fit model, data, and residuals are shown in the left panel of Figure~\ref{fig:fig-7} and the best-fit parameters are listed in Table~\ref{table:table-2}. 

We followed similar procedure to fit the EPIC-pn data. First, we fitted the 2-5 keV band by absorbed power-law model and found the power-law photon index to be $2.21\pm0.03$ which is consistent with the photon index obtained from the SXT data (within errors).  We then extrapolated down to 0.3 keV and also up to 10 keV. This shows the presence of a very strong soft excess below 2 keV and a broad emission line near 6 keV as shown in the right panel of Figure~\ref{fig:fig-6}. The observed soft excess was fitted with two blackbody components and the broad emission line was modelled with a Gaussian function. The broad emission line was centred at $\sim6.9$ keV and the width of the line was found to be greater than 0.8 keV. The model {\tt tbabs$\times$(blackbody1+blackbody2+powerlaw)} without any significant residuals resulted in $C/dof=189.1/171$. The best-fit model, data and residuals are shown in the right panel of Figure~\ref{fig:fig-7} and the best-fit parameters are listed in Table~\ref{table:table-2}.

As a strong soft X-ray excess is detected in the EPIC-pn data, the origin of this component would be important to investigate. To understand this, we plotted UVW1 and UVM2 data along with the X-ray data in 0.3-2 keV and 2-10 keV ranges, as shown in Figure~\ref{fig:fig-8}. Here, we used a time binsize of 500~s. It should however be noted that large time gaps in the data sets shown in Figure~\ref{fig:fig-8} prohibit quantitative estimation of any time lag between X-rays and UV emission. Nonetheless, the figure hints that the UV data sets appear to be delayed by $\sim$10 ks with respect to the X-ray emission. Such delay is expected in the X-ray reprocessing scenario and thus favours the blurred reflection for the origin of the soft excess.

\subsection{Short term variability: time resolved spectroscopy}

A clear positive trend in the hardness ratio and hard X-ray light curve from the SXT data is the first time finding in this AGN. Normally, this type of AGNs show negative correlation \citep{2009ApJ...705.1336C}. The observed positive correlation between the hard X-ray band and the hardness ratio is unique. This can be verified if we find some variations for power-law photon index with its flux. This can be done by using the time resolved spectroscopy technique. For this, we divided the entire SXT exposure into three parts with exposures of \ie~ $\sim~0-50$ks, $\sim~50-100$ ks and $\sim~100-170.5$ks (see vertical lines in the left graph of figure~\ref{fig:fig-3}). We then obtained the source spectra for all the three time windows and used the background spectrum and response matrices provided by the POC team. The effective area files (ARF) generated for the spectral fitting were applied here. We jointly fitted all three spectra using the best fit model obtained from the spectral modeling. We derived the parameters and unabsorbed fluxes in various bands as listed in Table~\ref{table:table-2}. From Table~\ref{table:table-2}, one can find a relation between power-law photon index ($\Gamma$) and power-law flux in the 0.3-10 keV band. It is clear that the power-law photon index increases with the decrease in power-law continuum flux (See bottom panel of Figure~\ref{fig:fig-9}). The best-fit model, data and residuals are shown in the top graph of Figure~\ref{fig:fig-9}.

 \section{Summary and Discussion}

We studied available long exposure X-ray observations (only 2) of NGC~4748 carried out in 2017 with {\it AstroSat} and in 2014 with {\it XMM-Newton}. Data obtained from both the observations show extreme variations in the soft X-ray as well as the hard X-ray bands. Along with large changes in the soft and hard X-ray bands, we found a soft X-ray excess in both the observations. SXT and EPIC-pn spectra were fitted by using phenomenological models such as power-law continuum and thermal blackbody emission. We found the following results from our analysis :

\begin{itemize}
\item A correlated variability of the soft and hard X-ray bands is seen.

\item The amplitude of variations in the hard X-ray band of SXT data is surprisingly higher, almost by a factor of forty-five which has never seen in this AGN. On the other hand, flux variations in the soft band in the SXT data varies by a factor of three and thirteen on short and long time scales, respectively.  For EPIC-pn data, both the soft and hard X-ray flux show similar variations (by a factor of two).

\item The changes seen in the soft and hard X-ray flux are positively correlated for both while positive and negative correlation between hard X-ray flux and hardness ratio were found for the 2017 and 2014 observations, respectively. The time-resolved spectroscopy confirms a ``harder when brighter" nature observed in 2017 data set.

\begin{figure*}
\includegraphics[scale=0.7]{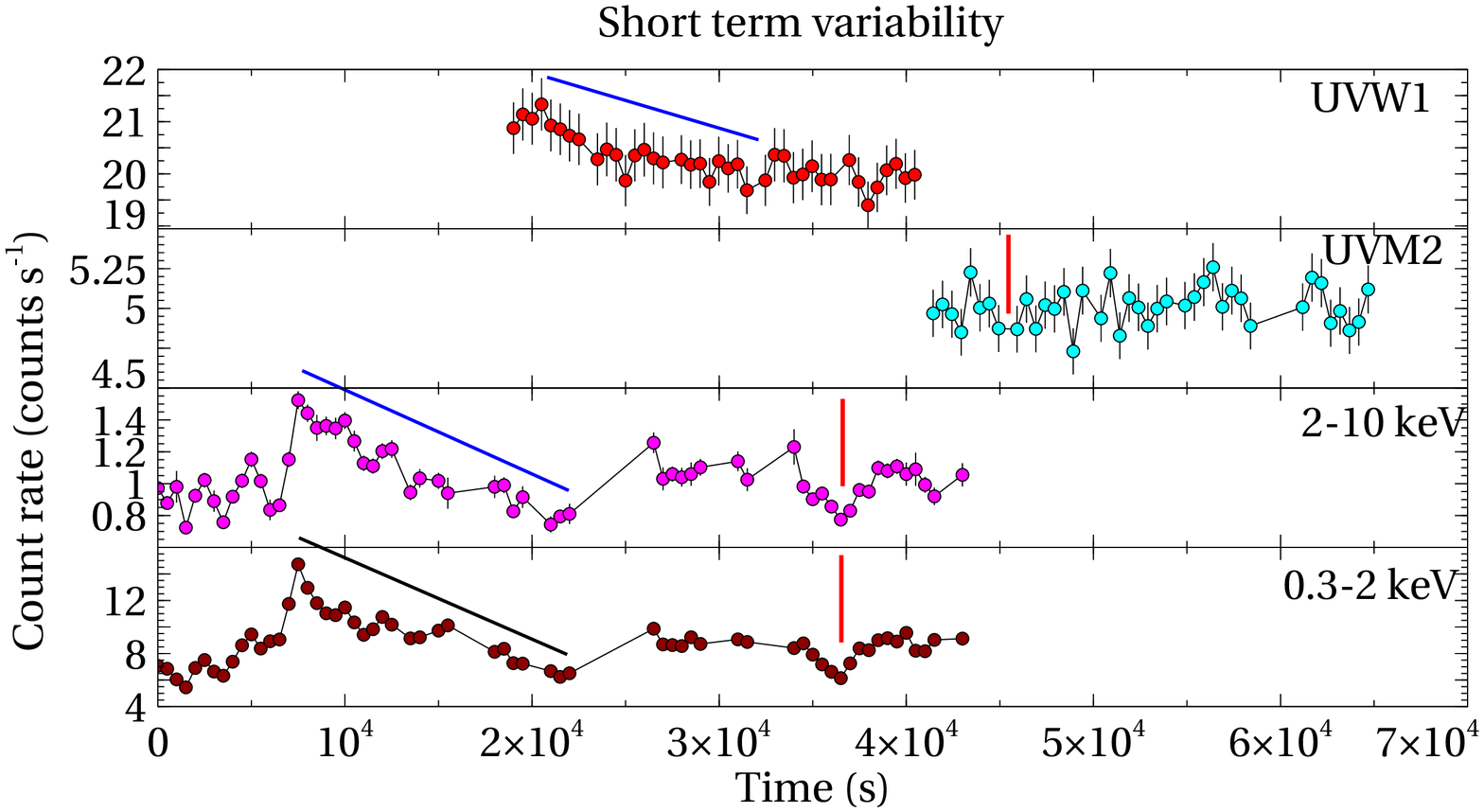}
\caption{500s binned light curves in (i) UVW1 and UVM2 bands obtained from Optical Monitor (ii) soft X-ray and hard X-ray bands extracted from EPIC-pn observations of NGC~4748. The straight lines hint the echoes of X-ray reprocessing process in the UV bands.}
\label{fig:fig-8}
\end{figure*}

\item Both the observations show the presence of soft X-ray excess below $\sim$2 keV over the power-law continuum. 
\item The observed soft excess in 2017 is well described by a single blackbody while the 2014 observation requires two blackbody components to fit the soft X-ray excess.
\item Average power-law continuum slopes are consistent (within error-bars) for both the observations, suggesting that the Comptonization may not be a dominant process during both the observations. However, a variation of photon index within the 2017 observation can be seen (see Figure 9). On the Other hand,  the 2014 data shows the presence of broad iron-K$\alpha$ emission line and hints a delayed UV emission. These findings do not favour a common origin of soft excess for both the data set.
\end{itemize}

The observed variations in the soft and hard X-ray bands and their correlations are interesting to understand the characteristics of NGC~4748. The variations seen in 2017 data set is different in both the soft X-ray and hard X-ray bands while the 2014 data shows similar changes in both the bands. The significant correlations in soft and hard X-ray bands suggest a relationship between these bands. Further, the positive offset in 2017 data suggests the presence of a slowly variable component in the soft X-ray emission while negative offset in the 2014 data set (loosely speaking) may infer a possible absorption in the soft band (see Figure~\ref{fig:fig-4}). However, we do not see any absorption signature in the soft X-ray band (see the residuals in the right panel of Figure~\ref{fig:fig-7}). The negative offset may be due to the presence of slowly variable spectral component such as broad Fe-K line along with a highly variable power-law continuum. \citet{2002MNRAS.335L...1F} studied a Seyfert~1 galaxy MCG~6-30-15 using about four days continuous {\it XMM-Newton} observation. They found that the skewed broad Fe-K emission line near 6 keV showed a minimum variability compared to any other spectral components in the X-ray band. As we detected a very broad iron line near 6 keV whose width is found to be greater than 0.8 keV, the negative offset is possibly caused due to a slowly varying broad iron emission line. Such a low variable broad iron emission line is to be originated by the fluorescence phenomena followed by photo-absorption of X-ray power-law emission in the partially ionized accretion disk near the supermassive black hole (\ie~\citet{2004MNRAS.349.1435M,2002MNRAS.335L...1F}).

Power-law continuum is believed to be the dominant spectral component in the energy band above 2 keV. Sometimes, the variations seen in hardness ratio depict the presence of the spectral features \ie,~neutral absorption along the line of sight or the changes in the geometry/size of the hot corona or the changes in the accretion flow state. The observed hardness ratio is highly variable and the correlation with hard X-ray flux appears to be strong in the 2017 data set while 2014 data shows moderate anti-correlation between the hardness ratio and hard X-ray band flux (see Figure~\ref{fig:fig-5}). The anti-correlation appears stronger when we used entire X-ray band with hardness ratio as shown in the bottom right panel of Figure~\ref{fig:fig-5}. Thus, the 2017 data set shows a ``harder when brighter" behaviour and the 2014 data set exhibits the ``softer when brighter" state of this AGN. From the Table-2, the power-law flux in the 0.3-10 keV band varies  $\sim30\%$ from 2014 to 2017 which appears to be a significant change likely associated with a change in the accretion rate. This type of a behaviour of NGC~4748 is interesting as this AGN has gone through likely a state transition from high flux state to low flux state during 2014-2017. Such state transition is analogous to the black hole binary systems \citep{2008ApJ...682..212W}. \citet{2008ApJ...682..212W} investigated the spectral evolution of six X-ray binary systems and found a critical transition point at Eddington accretion rate~$0.01$ and photon index $\Gamma\sim1.5$. These accretion states are normally known as low/hard and high/soft about this transition point and spectral states are completely different around such transition point. Typically in X-ray band, the low/hard state is described by a power-law continuum and the high/soft state is dominated by the thermal disk component along with power-law emission \citep{2006ARA&A..44...49R}. 

Similarly, \citet{2009MNRAS.399..349G} found a significant anti-correlation between the photon index and Eddington ratio for a number of low luminosity AGNs which was opposite to that seen in most of the AGNs. Recently a similar study was performed using \swift{}~data by \citet{2016MNRAS.459.3963C} and they found a ``harder when brighter'' trend in the low luminosity AGNs also. In the low luminosity AGNs, this is expected due to the radiatively inefficient accretion flow such as the advection dominated accretion flow \ie, \citep{1997ApJ...489..865E}. \citet{2006ApJ...646L..29S} studied a sample of radio-quiet AGNs and found that the power-law photon index depends primarily on the accretion rate expected in the geometrically thin and optically thick standard disk. Thus, normally radio-quiet AGNs show the positive correlation between $\Gamma$ and Eddington ratio being in ``softer when brighter" nature. Here, NGC~4748, being a radio quiet AGN, showed both ``harder when brighter" and ``softer when brighter" nature which seems rare in AGNs. Therefore, this AGN exhibits two states within a three year span likely with two different spectral states. 

\begin{figure}[!ht]
\begin{center}
\includegraphics[scale=0.3, angle=-90]{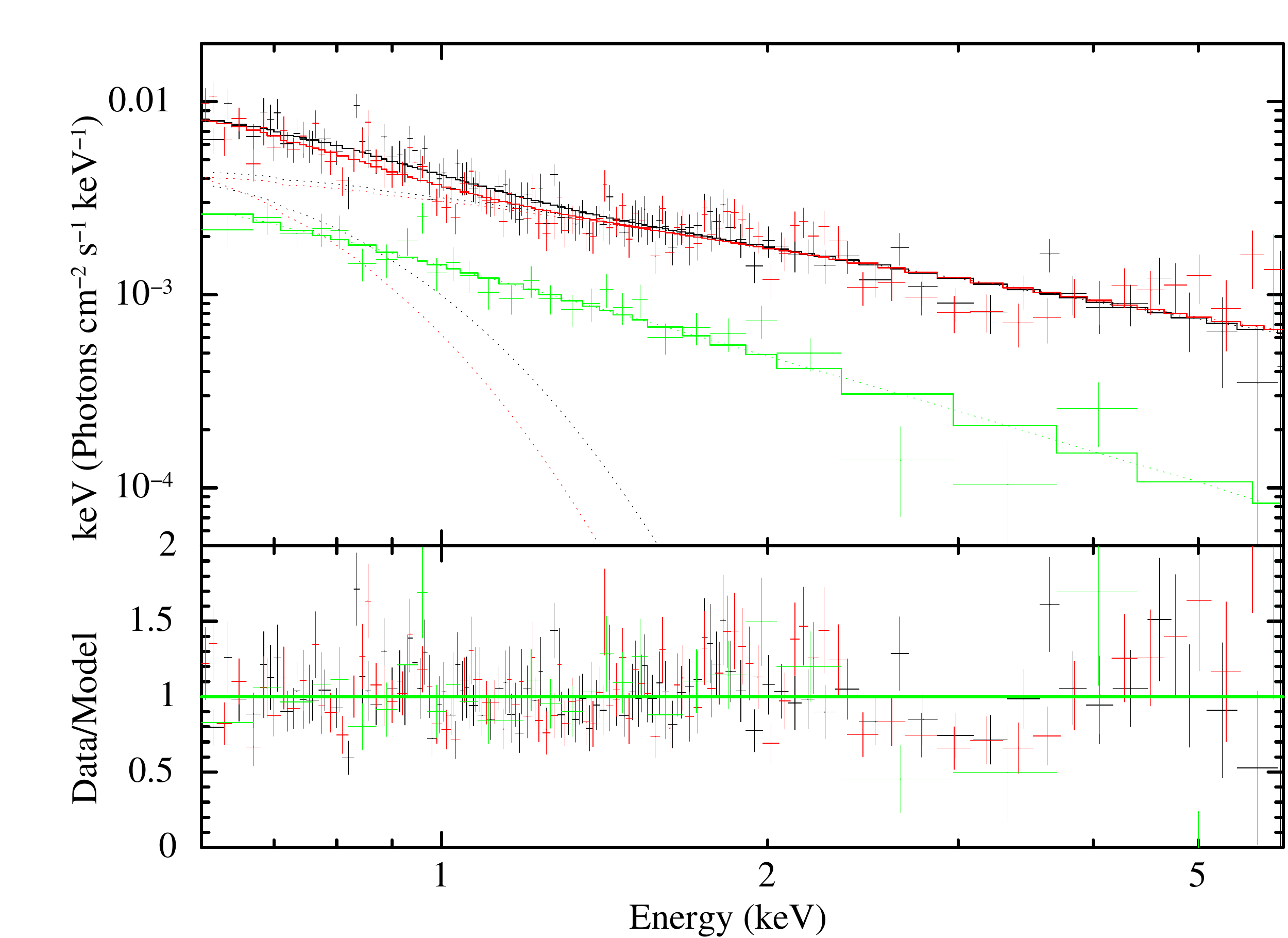}
\includegraphics[scale=0.5, angle=0]{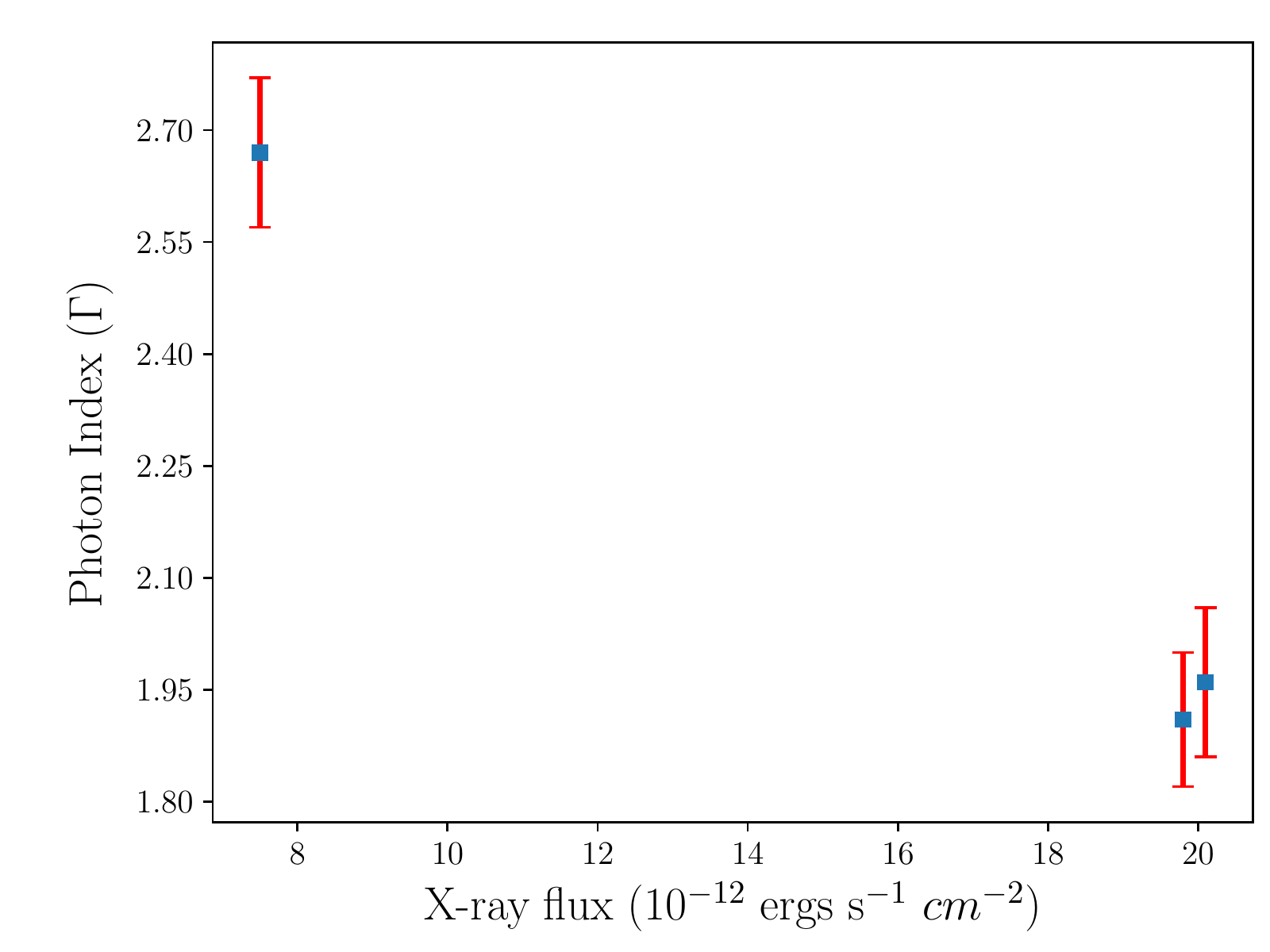}
\end{center}
\caption{\textbf{Top} Spectral fitting of data during all three segments of SXT exposures. The black, red and green colours represent the data and best-fit models for each segment. Dotted straight and sloppy lines are the power-law continuum and dotted curves show a blackbody model used to fit the soft X-ray excess. \textbf{Bottom}: An anti-correlation between photon index and the 0.3-10 keV power-law continuum flux is shown.}
\label{fig:fig-9}
\end{figure}

In the ``harder when brighter" state in 2017 data, the observed soft X-ray excess was described by a single blackbody with inner disk temperature of about 120 eV. This huge temperature is not acceptable theoretically and can be achieved by a physical phenomenon such as Comptonization process. \citet{2012MNRAS.420.1848D} described the soft X-ray excess as a result of cool Comptonization process. In this process, the seed photons from the disk are inverse Compton scattered in the optically thick (optical depth $>1$) and cool plasma of about 0.2 keV. This explanation is supported as we obtained a varying photon index during the 2017 observation. To understand this behaviour in more detail, we have proposed new long {\it AstroSat} observation which will be performed simultaneously in UV and X-ray from the onboard UVIT and SXT instruments.  

The strong soft X-ray excess observed in the 2014 data, on the other hand, may be associated with a different mechanism as it belongs to the ``softer when brighter" state of the AGN, a state similar to high/soft behavior in high accretion flow regime. The soft excess was modelled by two blackbody components with temperatures of $\sim80$ eV and $\sim180$ eV. To investigate its origin in the 2014 data, we found possibly a delayed UV emission of about 10 ks in the UVW1 ($\sim$2910 ~angstrom) and UVM2 ($\sim$2310~angstrom) bands (see Figure~\ref{fig:fig-8}). This delayed emission could be due to the X-ray reprocessing in the accretion disk. To confirm this, we used temperature profile obtained from the standard accretion disk \citep{1973A&A....24..337S} to estimate the light travel time for X-ray emission. Using its known mass $\sim2.6\times10^6$ from literature \citep{2009ApJ...705..199B} and acceptable Eddington ratio ($\sim0.01$), we obtained a time delay of $\sim10$ ks for UVW1 band. This is consistent with the time-delay observed in UVW1 band within an observational cadence. The trend and dips seen in the X-ray bands near 36 ks seem to appear in the UVM2 band near 45 ks shown in Figure~\ref{fig:fig-8}. These trend and dips support the scenario for the X-ray reprocessing on the outer disk. In addition, a broad Fe-K emission line seen around 6 keV suggests that this is originated from the vicinity of the SMBH and is blurred likely due to relativistic effects. The presence of strong and broad iron line near 6 keV and the delayed UV emission suggests that the origin of the soft X-ray excess is likely due to blurred reflection as a result of the strong light bending close to SMBH.          

\section*{Acknowledgements}
We are thankful to the anonymous referee for his/her useful comments which have improved the manuscript significantly. 
 MP is thankful for financial support from the UGC, India through the Dr. D. S. Kothari Post-Doctoral  Fellowship (DSKPDF) program (grant No. BSR/2017-2018/PH/0111). PK acknowledges support from the Aryabhatta Post-Doctoral Fellowship (A-PDF) grant (AO/A-PDF/770). This publication uses the data from {\it AstroSat} mission of the ISRO, archived at the Indian Space Science Data Centre (ISSDC). We thank members of the SXT team for their contribution to the development of the instrument and analysis software. We also acknowledge the contributions of the {\it AstroSat} project team at ISAC and IUCAA. This work has been performed utilising the calibration data-bases and auxiliary analysis tools developed, maintained and distributed by {\it AstroSat}/SXT team with members from various institutions in India and abroad. This research has also used data from the {\it XMM-Newton}, operated by the European Space Agency.

\vspace{-1em}

\newcommand{\pasp}{PASP} \def\apj{ApJ} \def\mnras{MNRAS}
\def\aap{A\&A} \def\apjl{ApJ} \def\aj{aj} \def\physrep{PhR}
\def\pre{PhRvE} \def\apjs{ApJS} \def\pasa{PASA} \def\apss{Ap\&SS} \def\pasj{PASJ}
\def\nat{Nat} \def\ssr{SSRv} \def\aapr{AAPR} \def\araa{ARAA}

\bibliographystyle{apj} 
\bibliography{refs}

\end{document}